# Improving Functional Connectome Fingerprinting with Degree-Normalization

Benjamin Chiêm [1,2], Kausar Abbas [3,4], Enrico Amico [5,6],
Duy Anh Duong-Tran [3,4], Frédéric Crevecoeur [1,2], Joaquín Goñi [3,4,7,*]

[1] Institute of Communication Technologies, Electronics and Applied Mathematics, Université catholique de Louvain, Louvain-la-Neuve, Belgium
[2] Institute of Neurosciences, Université catholique de Louvain, Louvain-la-Neuve, Belgium
[3] Purdue Institute for Integrative Neuroscience, Purdue University, West Lafayette, IN, USA
[4] School of Industrial Engineering, Purdue University, West Lafayette, IN, USA
[5] Institute of Bioengineering, Center for Neuroprosthetics, Ecole Polytechnique Fédérale de Lausanne, Lausanne, Switzerland
[6] Department of Radiology and Medical Informatics, University of Geneva, Geneva, Switzerland
[7] Weldon School of Biomedical Engineering, Purdue University, West Lafayette, IN, USA

* Corresponding author

## Abstract

Functional connectivity quantifies the statistical dependencies between the activity of brain regions, measured using neuroimaging data such as functional MRI BOLD time series. The network representation of functional connectivity, called a Functional Connectome (FC), has been shown to contain an individual fingerprint allowing participants identification across consecutive testing sessions. Recently, researchers have focused on the extraction of these fingerprints, with potential applications in personalized medicine.

Here, we show that a mathematical operation denominated degree-normalization can improve the extraction of FC fingerprints. Degree-normalization has the effect of reducing the excessive influence of strongly connected brain areas in the whole-brain network. We adopt the differential identifiability framework and apply it to both original and degree-normalized FCs of 409 individuals from the Human Connectome Project, in resting-state and 7 fMRI tasks.

Our results indicate that degree-normalization systematically improves three fingerprinting metrics, namely differential identifiability, identification rate and matching rate. Moreover, the results related to the matching rate metric suggest that individual fingerprints are embedded in a low-dimensional space.

The results suggest that low-dimensional functional fingerprints lie in part in weakly connected subnetworks of the brain, and that degree-normalization helps uncovering them. This work introduces a simple mathematical operation that could lead to significant improvements in future FCs fingerprinting studies.

## Impact statement

We introduce a simple mathematical operation that systematically improves the extraction of functional connectivity fingerprints from neuroimaging data, according to three different metrics. The results suggest that the information related to individual traits lies in part in weakly connected brain areas and can be compressed in a low-dimensional space. We also show the benefits of using multiple metrics to quantify fingerprint in a dataset. Our approach could improve future individual-level studies of functional neuroimaging data, which are crucial for the personalized diagnosis and treatment of neurological disorders, as well as for the study of the relationship between brain and behavior.

# 1. Introduction

The study of brain functional connectivity aims to understand how distributed neural Regions of Interest (ROIs) interact with each other during resting-state and task conditions [Bullmore and Sporns (2009); Fornito et al. (2016)]. Thanks to advances in functional Magnetic Resonance Imaging (fMRI), the measurement of Blood-Oxygenation-Level Dependent (BOLD) signals provides an estimate of brain activity across conditions [Ogawa et al. (1990)]. In this context, a widespread approach to quantify functional connectivity is to compute pairwise Pearson's correlation coefficients between BOLD time series measured at each ROI. The resulting symmetric correlation matrix is referred to as a Functional Connectome (FC) and can be understood as the adjacency matrix of a network where nodes are ROIs and edges represent functional interactions between those ROIs [Bullmore and Sporns (2009); Fornito et al. (2016)].

The network analysis of brain connectivity is able to capture important features of cortical organization, such as integration and segregation [Bullmore and Sporns (2009); Shine et al. (2016); Shine et al. (2018); Shine et al. (2019)], as well as modularity and community structure [Sporns and Betzel (2016); Betzel et al. (2016); Betzel et al. (2019); Puxeddu et al. (2020)]. Furthermore, FCs have been used in the study of several brain disorders [Fornito et al. (2015)] such as schizophrenia [Micheloyannis et al. (2006); Lynall et al. (2010); Gutiérrez-Gómez et al. (2020)] and Alzheimer disease [Supekar et al. (2008); Svaldi et al. (2019)]. Several studies demonstrated the existence of a fingerprint embedded in individual-level FCs, allowing participant identification in test-retest settings [Finn et al. (2015); Gratton et al. (2018); Mars et al. (2018); Satterthwaite et al. (2018); Pallarés et al. (2018); Liu et al. (2018); Iturria-Medina et al. (2018); Seitzman et al. (2019); Menon and Krishnamurthy (2019)]. This fingerprint can be extracted through data driven procedures [Amico and Goñi (2018)] as well as reproduced across sites [Bari et al. (2019)]. These findings have important implications in the perspective of individual-level functional connectivity analysis. For instance, personalized medicine in the study of brain disorders can benefit from FCs revealing robust individual traits [Iturria-Medina et al. (2018); Svaldi et al. (2019)].

As recently shown [Rajapandian et al. (2020)], fingerprints are also reflected in different network properties of FCs. To characterize the topology of functional networks, numerous networks statistics have been introduced [Rubinov and Sporns (2010)]. One of the most fundamental measure for binary networks is the degree of a node, i.e. the number of nodes it is connected to. In weighted networks, the weighted degree (or the strength) of a node is the sum of the weights of its neighboring edges. The weighted degree sequence denotes the vector gathering the weighted degree of all nodes in the network.

Here, we show the benefits of applying a mathematical operation, known as *degree-normalization*, to FCs prior to extracting functional connectivity fingerprints. Degree-normalization uses the information encoded in the weighted degree sequence in order to reduce the weight of edges lying between strongly connected nodes (hubs) comparatively to others, thereby balancing their excessive influence in the network. This operation has been applied in previous studies on weighted communicability measures of networks [Crofts and Higham (2009); Estrada et al. (2012); Rajapandian et al. (2020)] as well as in the study of random walks on networks through the use of the normalized Laplacian [Lambiotte et al. (2014)]. We adopt the differential identifiability framework recently developed by Amico and Goñi for FC fingerprinting [Amico and Goñi (2018)] based on a Principal Components Analysis (PCA) decomposition-reconstruction procedure. Because computing the absolute value of FCs is an intermediate step required prior to applying degree-normalization, we compare the results of this framework applied on (i) the original (signed) FCs, (ii) the FCs taken in absolute value and (iii) the degree-normalized FCs. In order to assess the quality of the fingerprint extraction, we consider two previously introduced metrics, namely differential identifiability [Amico and Goñi (2018)] and identification rate [Finn et al. (2015)], and we introduce a variant of the latter called matching rate. Our results show that degree-normalization improves the fingerprinting scores for all metrics and that reconstructing the corresponding optimally identifiable FCs requires fewer principal components compared to original FCs. We also highlight the difference in the interpretation of the identification rate and the matching rate and argue that the latter provides a more robust depiction of the individual fingerprint in FCs.

## 2. Materials and Methods

*2.1. Dataset*

We included 409 unrelated individuals from the Human Connectome Project (HCP) 1200-participants release [Essen et al. (2013)]. This subset of unrelated individuals was chosen from the overall dataset to ensure that no two participants have a shared parent. The criterion to exclude siblings (whether they share one or both parents) was crucial to avoid confounding effects in our analyses due to family-structure. Data from resting-state (REST) and seven functional Magnetic Resonance Imaging (fMRI) tasks were used: emotion processing, gambling, language, motor, relational processing, social cognition and working-memory. In this study, we will collectively refer to the resting-state and all the tasks as *conditions.*

For each condition, subjects underwent two sessions corresponding to two different phase-encoding directions (left-to-right and right-to-left). The resting-state fMRI scans were acquired on two different days with a total of four sessions (coded as REST1 and REST2). In this study, we used the two sessions from REST1. The HCP scanning protocol was approved by the Institutional Review Board at Washington University in St. Louis. Full details on the HCP dataset have been published previously [Essen et al. (2012); Glasser et al. (2013); Smith et al. (2013)].

The brain atlas used in this study is the multimodal parcellation MMP1.0 proposed by Glasser et al. [Glasser et al. (2016)] and comprising 180 cortical regions by hemisphere. For completeness, we added 14 subcortical regions (covering the bilateral striatum, thalamus, hippocampus and amygdala) provided by the HCP release, for a total of $N = 374$ Regions of Interest (ROIs).

*2.2. Preprocessing*

We used the *minimally preprocessed data* provided by the HCP [Glasser et al. (2013)]. This pipeline includes artifacts removal, motion correction, and registration to standard template. Full details on this pipeline can be found in earlier publications [Glasser et al. (2013); Smith et al. (2013)].

In addition, we applied the following processing steps to the extracted BOLD signals. For resting-state fMRI data: (i) we regressed out the global gray-matter signal from the voxel time courses [Power et al. (2014)], (ii) we applied a bandpass first-order Butterworth filter in the forward and reverse directions (0.001Hz to 0.08Hz ; Python function filtfilt from the Scipy package v1.2.1), and (iii) the voxel time courses were z-scored and then averaged per brain region, excluding any outlier time points that were outside of 3 standard deviations from the mean (Workbench software, command -cifti-parcellate). For task fMRI data, we applied the same steps, with a more liberal frequency range for the band-pass filter (0.001Hz to 0.25Hz) since the relationship between different tasks and optimal frequency ranges is still unclear [Cole et al. (2014)].

*2.3. Degree-Normalization of a Functional Connectome*

We compute a functional connectivity matrix **FC** as the $N \times N$ matrix of pairwise, zero-lag Pearson's correlation coefficients between the $N$ regional BOLD time series :

$$\mathbf{FC} = [\mathbf{FC}_{ij}] \tag{1}$$

where $\mathbf{FC}_{ij} \in [-1,1]$ and $\mathbf{FC}_{ij} = \mathbf{FC}_{ji}$. Without loss of generality, we ignore self-loops in the functional network by setting $\mathbf{FC}_{ii} = 0$. This matrix, which we denote as the **Baseline FC**, can be directly treated as the adjacency matrix of a weighted, undirected and *signed* network, as done in previous fingerprinting studies [Finn et al. (2015); Amico and Goñi (2018)]. In the present work, we also consider the *unsigned* version in order to avoid the occurrence of complex numbers due to the degree-normalization (see below). This is done by taking the entry-wise absolute value of correlation coefficients in **FC**. We denote this as the **Absolute FC**, $|\mathbf{FC}|$, with all entries verifying $|\mathbf{FC}|_{ij} \in [0,1]$.

The degree $d_i$ of node $i$ of an unsigned network is defined as the sum of the weights of its neighboring edges :

$$d_i = \sum_{j=1}^{N} |\mathbf{FC}|_{ij} \tag{2}$$

The degree matrix **D** is the $N \times N$ matrix containing the degree sequence on its diagonal, and zeros elsewhere:

$$\mathbf{D}_{ii} = d_i \tag{3}$$
$$\mathbf{D}_{ij} = 0, \quad \forall i \neq j \tag{4}$$

The degree-normalization of $|\mathbf{FC}|$ is mathematically defined as follows:

$$\mathcal{FC} = \mathbf{D}^{-1/2} |\mathbf{FC}| \mathbf{D}^{-1/2} \tag{5}$$

The resulting matrix $\mathcal{FC}$ is symmetric and corresponds to the adjacency matrix of the **Normalized FC** [Crofts and Higham (2009); Estrada et al. (2012)] where any excessive influence of nodes has been modulated by their corresponding weighted degree. Figure 1 summarizes the degree-normalization procedure. It is worth noting that degree-normalization on signed networks would potentially involve negative node degrees (Equation 2), which would in turn generate complex entries in the normalized FCs (Equation 5). For this reason, we restrict our analysis to the degree-normalization of unsigned FCs, i.e. FCs taken in absolute value.

*2.4. Functional Connectomes Fingerprinting*

We analyze each fMRI condition separately. In order to quantify the variability of our results in the population, we use sampling without replacement. We generate 100 random subsamples out of the 409 individuals in the database to obtain 100 datasets containing $K = 327$ (80% of 409) different individuals. For each condition, the dataset is composed of $2K = 654$ FCs, i.e. two FCs per individual corresponding to the two fMRI phase-encoding directions. Thus, we have for each individual a test FC and a retest FC. In order to extract functional connectivity fingerprints from this dataset, we adopt the differential identifiability framework based on group-level Principal Components Analysis (PCA) [Amico and Goñi (2018)]. In summary, the procedure consists of vectorizing the upper-triangular part (excluding diagonal values) of all FCs in the dataset, and then gathering these vectors in a data matrix of $\frac{N(N-1)}{2}$ rows associated to FC entries, and $2K$ columns associated to test-retest scans of each individual. Following the PCA decomposition of this matrix, FCs are reconstructed using an incrementally increasing number of components, selected in decreasing order of explained variance.

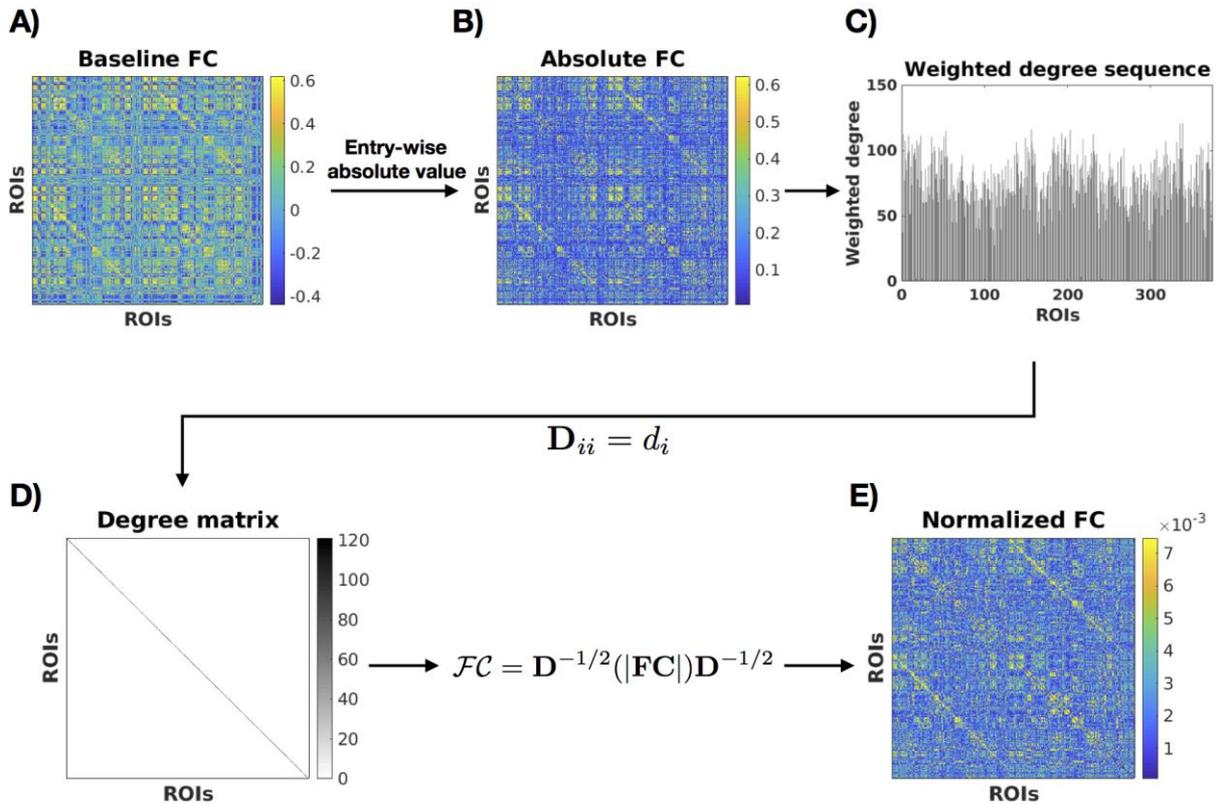

**Figure 1 : Degree-normalization of a Functional Connectome (FC)**. **A)** A functional connectome is computed as a matrix of pairwise Pearson's correlation coefficients between regional BOLD time series. Hence all values in the *Baseline FC* are within the range $[-1,1]$. **B)** The next step consists of taking the absolute value of all entries, which produces the *Absolute FC*, denoted by $|\mathbf{FC}|$. **C)** From that unsigned FC, we can extract the weighted degree sequence. **D)** The degree matrix $\mathbf{D}$ is a square matrix containing the weighted degree sequence on its diagonal and zeros elsewhere. **E)** Finally, we apply degree-normalization (Equation 5) to obtain the *Normalized FC*.

For each number of components, we compute the identifiability matrix $\mathbf{A} \in [-1,1]^{K \times K}$. The element $\mathbf{A}_{ij}$ is the entry-wise Pearson's correlation coefficient between the test FC of individual $i$ and the retest FC of individual $j$. Therefore, the diagonal elements $\mathbf{A}_{ii}$ represent the individuals' self-similarity between test and retest, while off-diagonal elements represent between-individuals similarities. Importantly, this means that $\mathbf{A}$ is not symmetric. Intuitively, the higher the contrast between diagonal and off-diagonal elements, the better are the extracted fingerprints.

## *2.5. Quantifying the Level of Identifiability*

We consider three metrics to estimate the amount of fingerprint in each subsample: the differential identifiability ($I_{\text{diff}}$), the identification rate ($ID_{\text{rate}}$) and the matching rate ($M_{\text{rate}}$). Let $I_{\text{self}} = \langle \mathbf{A}_{ii} \rangle$ denote the average of the diagonal elements of the identifiability matrix and let $I_{\text{others}} = \langle \mathbf{A}_{ij} \rangle$, $i \neq j$ be the average of the off-diagonal elements. The differential identifiability score ($I_{\text{diff}}$) [Amico and Goñi (2018)] is then defined as

$$I_{\text{diff}} = (I_{\text{self}} - I_{\text{others}}) * 100 \tag{6}$$

Each time a diagonal element $\mathbf{A}_{ii}$ is the highest of its row, we state that individual $i$'s retest FC has been correctly identified on the basis of his test FC. The identification rate [Finn et al. (2015)] is then

$$ID_{\text{rate}} = \frac{\text{Number of correctly identified individuals}}{\text{Total number of individuals}} \tag{7}$$

As we can also compute this metric column-wise (i.e. test FC identified from retest FC), we report the average of row-wise and column-wise $ID_{\text{rate}}$. Note that as per [Finn et al. (2015)], $ID_{\text{rate}}$ is a procedure with replacement, such that the algorithm was not forced to identify a unique subject on each iteration within a condition.

It might happen that the test FC of an individual $i$ is most similar not only to its own retest FC, but also to that of other individuals. In the extreme case of an FC being highly similar to many others, this will negatively impact the identification rate since many individuals will not be correctly identified. To remedy this, we propose a variant of identification rate, called matching rate ($M_{\text{rate}}$), where every time an FC from test session is matched with a retest FC (or vice versa) using the highest value of correlation along a row (or column) of an identifiability matrix, the matched test-retest pair is removed before the next comparison is made. In other words, $M_{\text{rate}}$ is equivalent to $ID_{\text{rate}}$ but *without* replacement. This way, all FCs are matched only once, no matter if they are similar to many others or not.

## 2.6. Control Experiment: Surrogate Degree-Normalization

In the present work, we evaluate the impact of normalizing each FC by its own degree sequence. As a control experiment, we also report the results of normalizing each FC by the degree sequence of a surrogate individual chosen uniformly at random, a process denoted as *surrogate degree-normalization*. Mathematically, this comes down to performing the fingerprinting analysis with the following normalized FCs for individual $u$ with surrogate $v$:

$$\mathcal{FC}_{u,\,\text{surr}} = \mathbf{D}_v^{-1/2} |\mathbf{FC}|_u \mathbf{D}_v^{-1/2} \tag{8}$$

Here, $|\mathbf{FC}|_u$ is the absolute FC of individual $u$, $\mathbf{D}_v$ is the degree matrix of individual $v$ and $\mathcal{FC}_{u,\,\text{surr}}$ is the surrogate-normalized FC of individual $u$. The operation is done for both test and retest FCs keeping the same surrogate individual. In the manuscript, normalizing an FC by its own degree sequence is sometimes refered to as *self degree-normalization* to avoid any ambiguity with surrogate degree-normalization.

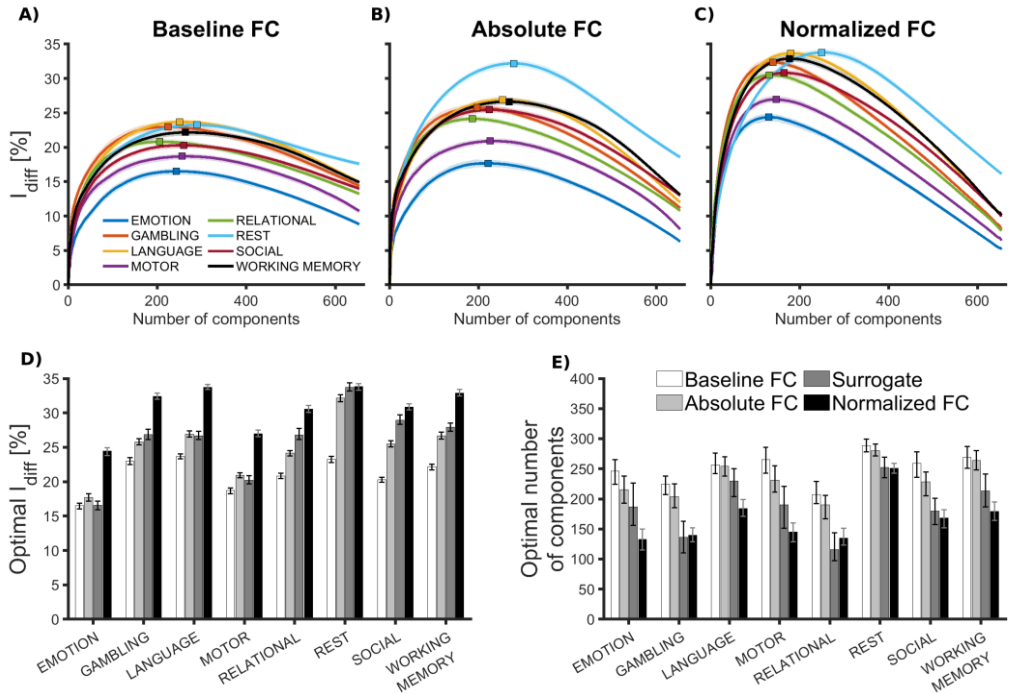

**Figure 2: Impact of degree-normalization on differential identifiability.** Panels **A)**, **B)** and **C)** present the evolution of $I_{\text{diff}}$ with respect to the number of principal components used for FCs reconstruction, for baseline, absolute and normalized FCs respectively. Solid lines represent the median value across 100 random subsamples (without replacement) of the database and shaded areas correspond to the inter-percentile range (2.5 and 97.5 percentiles). Square symbols highlight the optimum $I_{\text{diff}}$ of the median curves. **D)** Comparison of optimal $I_{\text{diff}}$ values for baseline, absolute, surrogate-normalized and self-normalized FCs. Error bars show the inter-percentile range (2.5 and 97.5 percentiles) across 100 random subsamples of the database. **E)** Number of principal components corresponding to the optimal $I_{\text{diff}}$ values of panel **D**.

# 3. Results

We apply the differential identifiability framework [Amico and Goñi (2018)] to baseline, absolute and normalized FCs. We compute three metrics : differential identifiability score ($I_{\text{diff}}$) [Amico and Goñi (2018)], identification rate ($ID_{\text{rate}}$) [Finn et al. (2015)] and the newly introduced matching rate ($M_{\text{rate}}$). The analysis is done for each fMRI condition separately, and performed independently on the 100 randomly drawn subsamples.

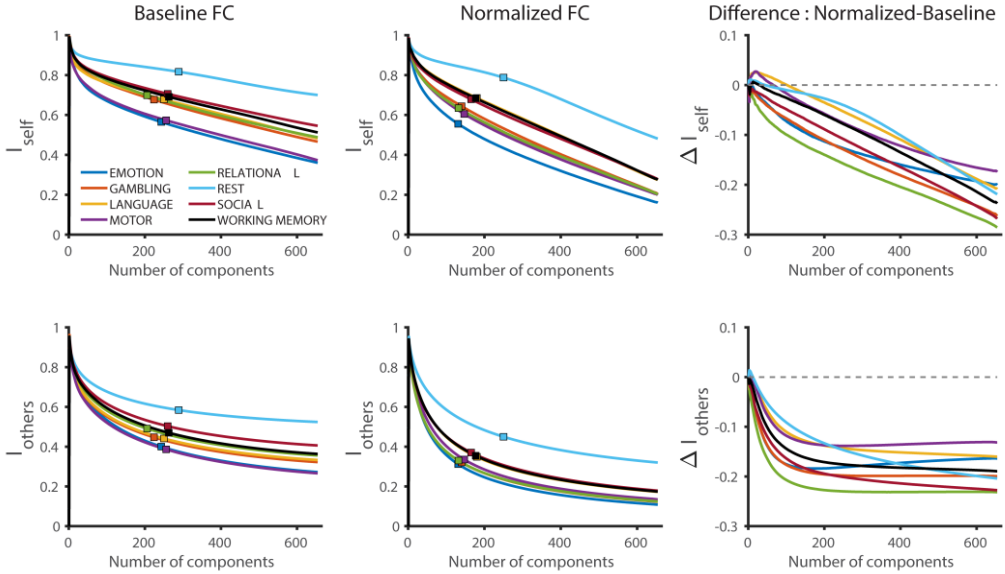

**Figure 3: Impact of degree-normalization on $I_{\text{self}}$ and $I_{\text{others}}$.** Top row : evolution of $I_{\text{self}}$ with the number of principal components added in descending order of explained variance for Baseline FC (left) and Normalized FC (middle). Top right shows $\Delta I_{\text{self}}$, which is the pointwise $I_{\text{self}}$ difference between Baseline FC and Normalized FC along principal components. Bottom row shows the analogous analyses for $I_{\text{others}}$, including $\Delta I_{\text{others}}$ at the bottom right. Optimal number of components for maximizing $I_{\text{diff}}$ are shown as square symbols in all cases.

Figure 2 presents the results related to differential identifiability ($I_{\text{diff}}$). We observe that the evolution of $I_{\text{diff}}$ with respect to the number of principal components used for FCs reconstruction is concave, with sharper curves in the case of normalized FCs, for all fMRI conditions. Figure 2D compares the optimal value of differential identifiability reached for baseline, absolute, surrogate-normalized and self-normalized FCs. We see that absolute and surrogate-normalized FCs achieve better scores than baselines FCs, for all conditions except the emotion processing task. Self-normalized FCs provide the best $I_{\text{diff}}$ scores for all fMRI conditions, with an average gain of 9.6% between baseline and self-normalized FCs (minimum gain: 7.9% for emotion ; maximum gain: 10.73% for working memory). We notice that in resting-state, surrogate degree-normalization leads to results that are comparable to that of self degree-normalization. Figure 2E shows the number of principal components corresponding to the optimal $I_{\text{diff}}$ values of Figure 2D. We observe that absolute and surrogate-normalized FCs require fewer components than baseline FCs, for all conditions except the language processing task and the working memory task for which baseline FCs and absolute FCs require a similar number of components. Self-normalized FCs require the lowest number of components, except for the gambling task and the relational processing task for which surrogate and self degree-normalization require a comparable number of components.

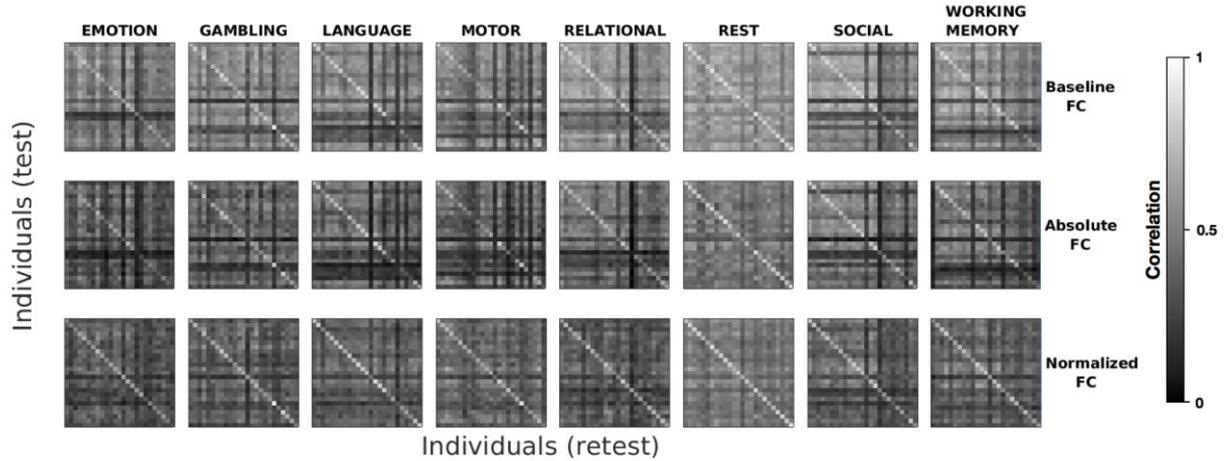

**Figure 4: Impact of degree-normalization on identifiability matrices.** Top row: Identifiability matrices obtained with the baselines FCs at the optimal $I_{\text{diff}}$ value, for all fMRI conditions. For visualization purposes, only 25 randomly selected individuals of one subsample of the database are displayed. Middle and bottom rows show the same analysis for absolute and self-normalized FCs respectively.

Figure 3 reports the behavior of $I_{\text{self}}$ and $I_{\text{others}}$. Overall, both metrics decrease with the number of principal components kept for FCs reconstruction. However, we observe for normalized FCs that $I_{\text{others}}$ decreases faster than $I_{\text{self}}$ in the first 200 components. This observation is valid for all fMRI conditions. In Figure 4, we display identifiability matrices obtained with baseline, absolute and normalized FCs, at the optimal $I_{\text{diff}}$ reconstruction point. We observe that diagonal elements stand out in all cases, indicating that individuals' self-similarity is correctly captured. Moreover, we see that degree-normalization smooths the distribution of off-diagonal elements while maintaining a good contrast with diagonal elements. Figure 5 highlights, for the motor task as an example, how degree-normalization is able to correct the identifiability profile of some individuals. This observation is valid for all fMRI conditions (results not shown).

Figure 6 presents the results related to the identification rate ($ID_{\text{rate}}$). Overall, the $ID_{\text{rate}}$ curves are also concave with a sudden rise in the last 50 components, for all fMRI conditions. This phenomenon is particularly pronounced for normalized FCs and highlights a shortcoming of the identification rate metric. As shown in Supplementary Figure S1, the identification rate is driven down by a few FCs being highly similar to others when around 600 principal components out of 654 are used for reconstruction. The last components then correct this bias. Figure 6D compares the optimal identification rates reached for baseline, absolute, surrogate-normalized and self-normalized FCs. We see that baseline and absolute FCs provide comparable results, while surrogate-normalization lowers the identification with respect to baseline FCs, for all conditions. Self-normalized FCs provide the best identification rates for all conditions, with an average gain of 16% with respect to baseline FCs (minimum gain: 6% for resting-state ; maximum gain: 30% for the motor task). Figure 6E shows the number of principal components corresponding to the optimal identification rates of Figure 6D. We see that self-normalized FCs require the lowest number of components, for all fMRI conditions. We observe large error bars (2.5-97.5 inter-percentile range across 100 random subsamples) in the case of surrogate-normalized FCs for the gambling task, the motor task and the working memory task. This come from the fact that in the realization of sampling without replacement, the highest $ID_\text{rate}$ is sometimes reached using all the components and sometimes with around 200 components, leading to a bimodal distribution of the optimal number of components. Ultimately, this produces large error bars. This phenomenon occurs particularly often with surrogate degree-normalization.

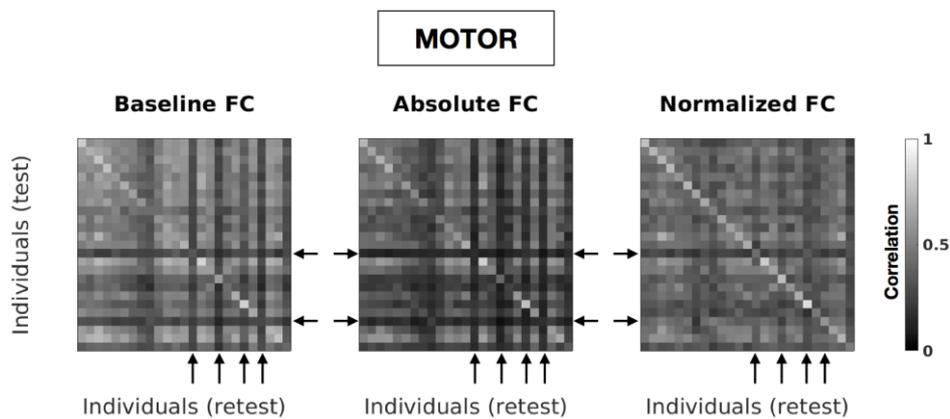

**Figure 5: Degree-normalization corrects the profile of outlier FCs.** Zoom on the panels of Figure 4 related to the motor task. Arrows highlight typical examples of FCs that are very different to any other FC in the cohort. Note that this effect is alleviated after degree-normalization.

Figure 7 presents the results related to the matching rate ($M_\text{rate}$). The $M_\text{rate}$ curves increase quickly until they reach a plateau value, except for the emotion processing and the motor tasks with baseline and absolute FCs. Importantly, the sudden rise in the last few components observed with identification rate does not occur with matching rate. Figure 7D compares the optimal matching rates reached for baseline, absolute, surrogate-normalized and self-normalized FCs. The observations made for identification rate are still valid for matching rate. Self-normalized FCs provide the best matching rates for all conditions, with an average gain of 14% with respect to baseline FCs (minimum gain: 5% for resting-state ; maximum gain: 22% for the motor task). Figure 7E shows the number of principal components corresponding to the values shown in Figure 7D. We see that normalized FCs require the lowest number of components, for all fMRI conditions. The large error bars (2.5 97.5 inter-percentile range across 100 random subsamples) for all conditions and all FCs are the result of the noisy plateau behavior of $M_{\text{rate}}$ curves. Indeed, depending on the subsample, the optimal matching rate can be achieved in a large range of number of components although its actual value remains stable.

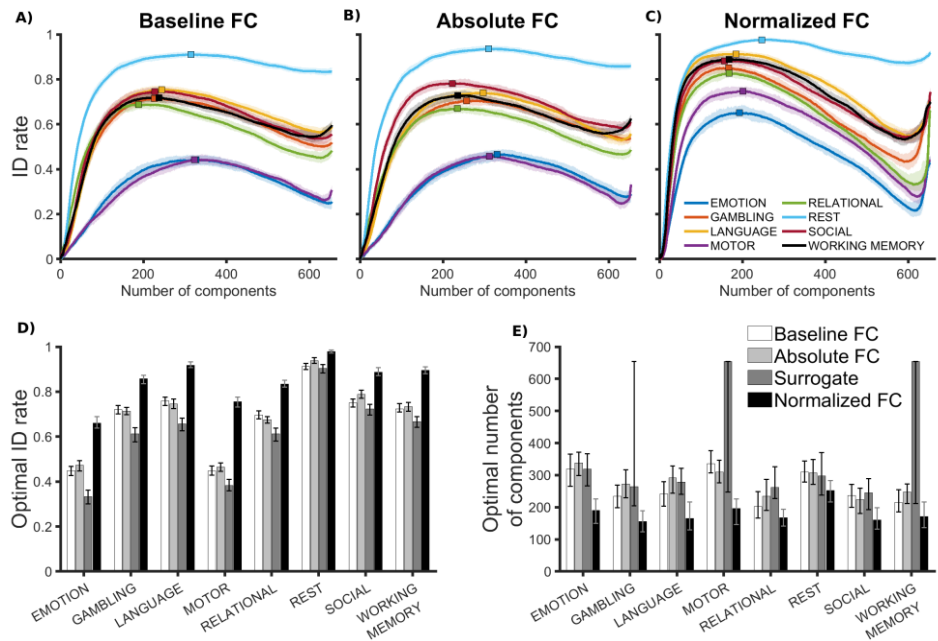

**Figure 6: Impact of degree-normalization on identification rate.** Panels **A)**, **B)** and **C)** present the evolution of $ID_{\text{rate}}$ with respect to the number of principal components used for FCs reconstruction, for baseline, absolute and normalized FCs respectively. Solid lines represent the median value across 100 random subsamples of the database and shaded areas correspond to the inter-percentile range (2.5 and 97.5 percentiles). Square symbols highlight the optimum $ID_{\text{rate}}$ of median curves. **D)** Comparison of optimal identification rates for baseline, absolute, surrogate-normalized and normalized FCs. Error bars show the inter-percentile range (2.5 and 97.5 percentiles) across 100 random subsamples of the database. **E)** Number of principal components corresponding to the optimal identification rates of panel **D**.

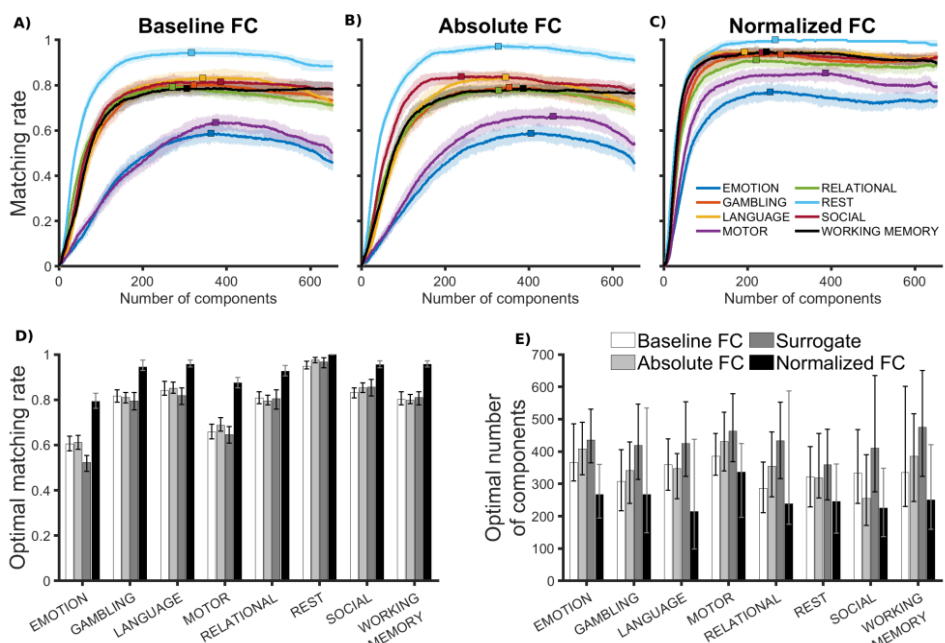

**Figure 7: Impact of degree-normalization on matching rate.** Panels **A)**, **B)** and **C)** present the evolution of $M_{\text{rate}}$ with respect to the number of PCA components used for FCs reconstruction, for baseline, absolute and normalized FCs respectively. Solid lines represent the median value across 100 random subsamples of the database and shaded areas correspond to the inter-percentile range (2.5 and 97.5 percentiles). Square symbols highlight the optimum $M_{rate}$ of median curves. **D)** Comparison of optimal matching rates for baseline, absolute, surrogate-normalized and normalized FCs. Error bars show the inter-percentile range (2.5 and 97.5 percentiles) across 100 random subsamples of the database. **E)** Number of PCA components corresponding to the optimal matching rates of panel **D**.

## 4. Discussion

Extracting fingerprints from Functional Connectomes (FCs) is an important challenge for future individual-level studies of functional connectivity. Here, we showed that the degree-normalization of FCs improves the fingerprinting process, according to three different metrics: differential identifiability, identification rate and matching rate. Moreover, the results indicate that the fingerprint of degree-normalized FCs is embedded in a lower-dimensional space (and hence can be compressed), compared to baseline FCs.

### *4.1. Improved Fingerprinting in a Lower-Dimensional Space*

Throughout our results, we observed that normalizing FCs improves the three fingerprinting scores considered in this work (Figures 2D, 6D and 7D), for all fMRI conditions. Moreover, these scores are achieved with fewer principal components than in the baseline and absolute cases (Figures 2E, 6E and 7E). This suggests that the degree-normalization reduces the individuals' fingerprints to a first set of principal components (in descending order of explained variance). In addition, when looking at the cumulative percentage of explained variance of the principal components extracted from the dataset (Supplementary Figure S2), we observe a reduced dominance effect. In other words the individual contribution of components to the explained variance is much more homogeneous. Together, these results indicate that the variance preserved by the components of normalized FCs, although lower, is highly specific to the contrast between individuals. From this perspective, degree-normalization could be beneficial for future FCs fingerprinting research.

### *4.2. Surrogate Degree-Normalization Improves Differential Identifiability, but not Identification Rate or Matching Rate*

Figure 2 shows that differential identifiability is improved following degree-normalization for several conditions, no matter if the correspondence between FCs and their respective degree sequence is preserved (self-normalization) or not (surrogate-normalization). Normalizing FCs has a global effect lowering the influence of hubs in the network [Crofts and Higham (2009); Estrada et al. (2012); Rajapandian et al. (2020)], which in turn allows better fingerprints to be extracted. This suggests that individual-specific components of functional connectivity might lie (in part) in sparsely connected areas whose contribution to the whole network is brought out by degree-normalization. The fact that surrogate degree-normalization sometimes improves differential identifiability compared to baseline indicates that the weighted degree sequence of FCs is similar across individuals. In Supplementary Figure S3, we show the results of the differential identifiability framework applied on degree sequences instead of functional connectivity matrices. We see that the weighted degree sequence alone imparts a moderate fingerprinting power no matter the number of components kept for the reconstruction, which was previously reported [Rajapandian et al. (2020)]. However, matching FCs with their own degree sequence for degree-normalization appears to be beneficial to all metrics and all fMRI conditions, while surrogate-normalization has a null or negative effect on the identification rate and the matching rate, compared to baseline (Figures 2D, 6D and 7D). This indicates that the normalization of FCs by their respective weighted degree sequence helps uncovering fingerprints and suggests a synergistic effect that goes beyond the fingerprints of original FCs and degree sequences separately.

*4.3. Matching Rate as a Correction of Identification Rate*

In this work, we observed that the identification rate metric, which has been used in several previous studies [Finn et al. (2015); Amico and Goñi (2018)], is sometimes driven down by a few individuals being highly similar to many others (Figure 6C and Supplementary Figure S1). Based on these results, it is noteworthy that the identification rate of an entire dataset can be compromised by a few or even one subject or single session of an otherwise high-quality fingerprinting dataset. In order to take into account the reality that each individual in our setting appears only once in each of the test and retest datasets, we introduced the matching rate metric. We noticed that the matching rate results are characterized by a plateau value (Figures 7A, B and C) rather than a concave behavior with a well-defined maximum, as obtained with differential identifiability and identification rate (Figures 2A, B, C and 6A, B, C). This suggests that, from the perspective of the matching rate metric, the PCA decomposition does not uncover functional connectivity fingerprints, but rather detects the dimensionality to which the data can be compressed while preserving an optimal fingerprinting power.

*4.4. Pros and Cons of Different Fingerprinting Metrics*

As discussed in section 4.2, while surrogate degree-normalization increases $I_{\text{diff}}$, its effects on $ID_{\text{rate}}$ and $M_{\text{rate}}$ are either neutral or negative, when compared to Baseline FCs. This highlights the limitations of $I_{\text{diff}}$ as a metric, where we see that even though surrogate degree-normalization has improved the overall contrast between self- and between-subject similarity (increased $I_{\text{diff}}$), its effects on the self-similarity are mostly negative (null or decreased $ID_{\text{rate}}$ and $M_{\text{rate}}$). On the other hand, we have discussed in section 4.3 the limitations of $ID_{\text{rate}}$ as a metric, where it can be severely affected by one or a few subjects/sessions of FCs that have high similarity with the rest of the population, hiding the underlying fingerprint of the dataset; this problem can be alleviated using $M_{\text{rate}}$ instead of $ID_{\text{rate}}$. At the same time, we observed (Figure 7) that $M_{\text{rate}}$ does not provide enough variation with number of principal components to find a clear optimal point of reconstruction in the differential identifiability frameworks. All these observations highlight that we should use more than one (preferably all three) metrics to estimate the amount of fingerprint in an FC data to avoid any unforeseen pitfalls. In other words, these three metrics represent a different face of the fingerprint in a sample of FCs.

## 4.5. Limitations and Future Work

The present work has several limitations. First, we chose to keep for each condition the total number of fMRI volumes available in the database. Previous work reported that larger numbers of frames can positively impact fingerprinting metrics [Amico and Goñi (2018); Abbas et al. (2020b)]. Here, as different scanning durations were used for each condition (see Supplementary Table T1), our results should be interpreted in light of this limitation. Future work should investigate whether degree-normalization is beneficial to fingerprinting studies using short scanning durations. Additionally, the effect of degree-normalization during functional reconfiguration could be assessed in scanning sessions that combine resting periods and tasks [Amico et al. (2020)].

Second, we conducted our experiments on a single dataset and used a particular brain parcellation. In order to evaluate the variability of our results with respect to variations in the dataset, we used sampling without replacement. Future work should reassess the impact of degree-normalization on external datasets, possibly obtained with different preprocessing pipelines [Parkes et al. (2018)]. We are confident that the results presented here are generalizable to other datasets and parcellations, since the other fingerprinting frameworks have been shown to reproducible across fMRI conditions [Finn et al.(2015)], robust across brain atlases [Amico and Goñi (2018); Abbas et al. (2020a)], and across scanning sites [Bari et al. (2019)]. Future work should include the assessment of this framework for studying brain injuries and neurological disorders.

Lastly, in the construction of the identifiability matrix, we considered the statistical similarity between reconstructed FCs, operationalized by the entry-wise Pearson's correlation coefficient. In contrast, recent studies recommended considering the geometric similarity of FCs, leveraging the observation that signed FCs lie on the manifold of positive semi-definite matrices and are therefore associated with a geodesic distance [Venkatesh et al. (2020); Abbas et al. (2020b)]. However, we would like to note that taking functional connectivity in absolute value, as required by the degree-normalization, breaks the positive semi-definiteness of FCs and therefore proscribes the geometric approach. Besides, the degree-normalization procedure is parameter-less whereas the geometric approach involves a dataset-dependent regularization parameter [Abbas et al. (2020b)]. Overall, we suggest that future work should consider statistical or geometric similarity depending on the context and application of the study.

## 5. Conclusion

Fingerprints extraction from Functional Connectomes (FCs) is an important step towards refined individual-level studies of brain connectivity, with potential applications in personalized medicine. In this report, we showed that the degree-normalization of FCs is a simple, parameter-less mathematical operation producing significant improvements of the fingerprinting quality, according to three different metrics, in resting-state and several task conditions. Furthermore, we argued that the fingerprint of FCs can be compressed in a low dimensional space, especially thanks to degree-normalization. We also show the potential benefits and pitfalls of three different fingerprinting metrics, where each of them uncovers different aspects of the fingerprint present in a sample of FCs. Overall, our results suggest that applying degree-normalization to FCs can be beneficial for future research focused on individual differences in brain networks.

## Acknowledgements

Benjamin Chiêm is a FRIA (F.R.S.-FNRS) fellow. The authors would like to thank Jean-Charles Delvenne for his helpful comments and suggestions.

Data were provided by the Human Connectome Project, WU-Minn Consortium (Principal Investigators: David Van Essen and Kamil Ugurbil; 1U54MH091657) funded by the 16 NIH Institutes and Centers that support the NIH Blueprint for Neuroscience Research; and by the McDonnell Center for Systems Neuroscience at Washington University

## Funding Statement

B.C. is a FRIA fellow (Grant N° 1.E051.18+F, Fonds pour la Formation à la Recherche dans l'Industrie et dans l'Agriculture, Fonds de la Recherche Scientifique, Belgium).

E.A. acknowledges financial support from the SNSF Ambizione project "Fingerprinting the brain: network science to extract features of cognition, behavior and dysfunction" (Grant N° PZ00P2_185716).

J.G. acknowledges financial support from NIH R01EB022574, NIH R01MH108467, Indiana Alcohol Research Center P60AA07611, and Purdue Discovery Park Data Science Award "Fingerprints of the Human Brain: A Data Science Perspective".

# Supplementary Materials

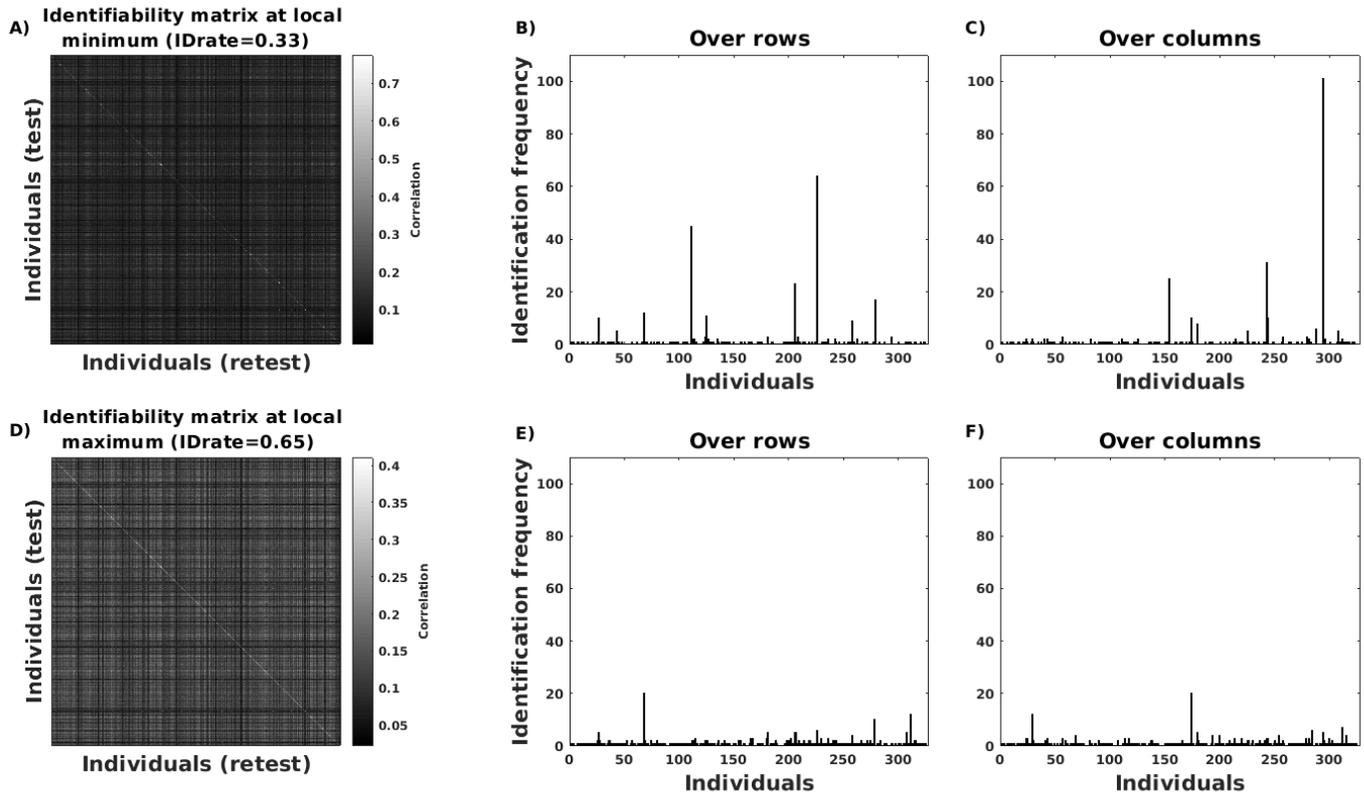

**Fig. S1. Identification rate is driven down by a few individuals.** We illustrate the undesired behavior of the identification rate ($ID_{rate}$), using one typical subsample of the database related to normalized FCs of the relational processing task (see Figure 6C). **A)** Identifiability matrix computed at a local minimum of the $ID_{rate}$ curve ($ID_{rate}$=0.33, 617 components). **B)** (resp. **C)**) Number of occurences of each individual as the maximum of the rows (resp. columns) of the identifiability matrix. The bottom row shows the same analysis when using all components ($ID_{rate}$=0.65, 654 components, right end of the $ID_{rate}$ curve in Figure 6C).

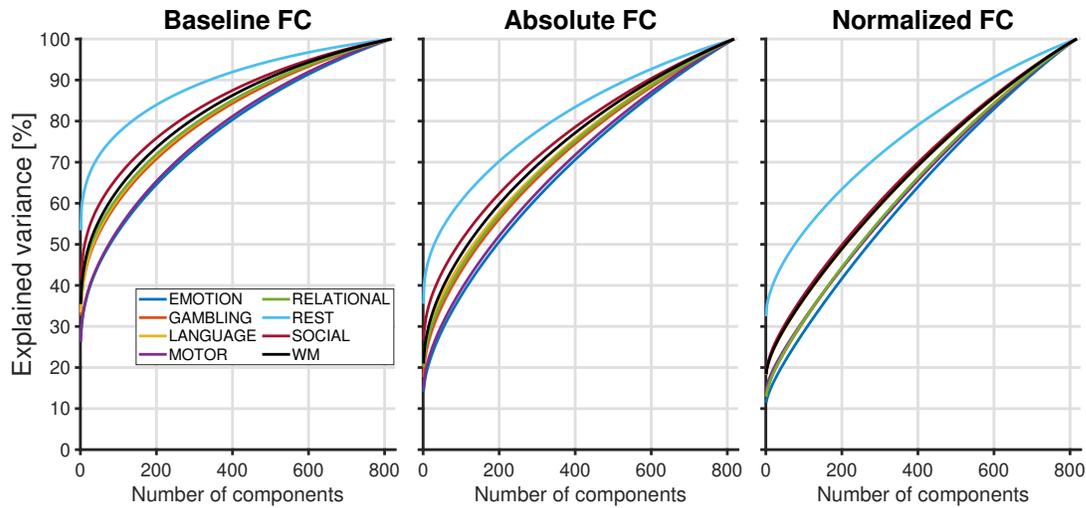

**Fig. S2. Cumulative percentage of explained variance.** Components resulting from the PCA decomposition of the database (409 individuals, 2 scans by individual) are added in decreasing order of explained variance, for baseline (left), absolute (middle) and normalized FCs (right).

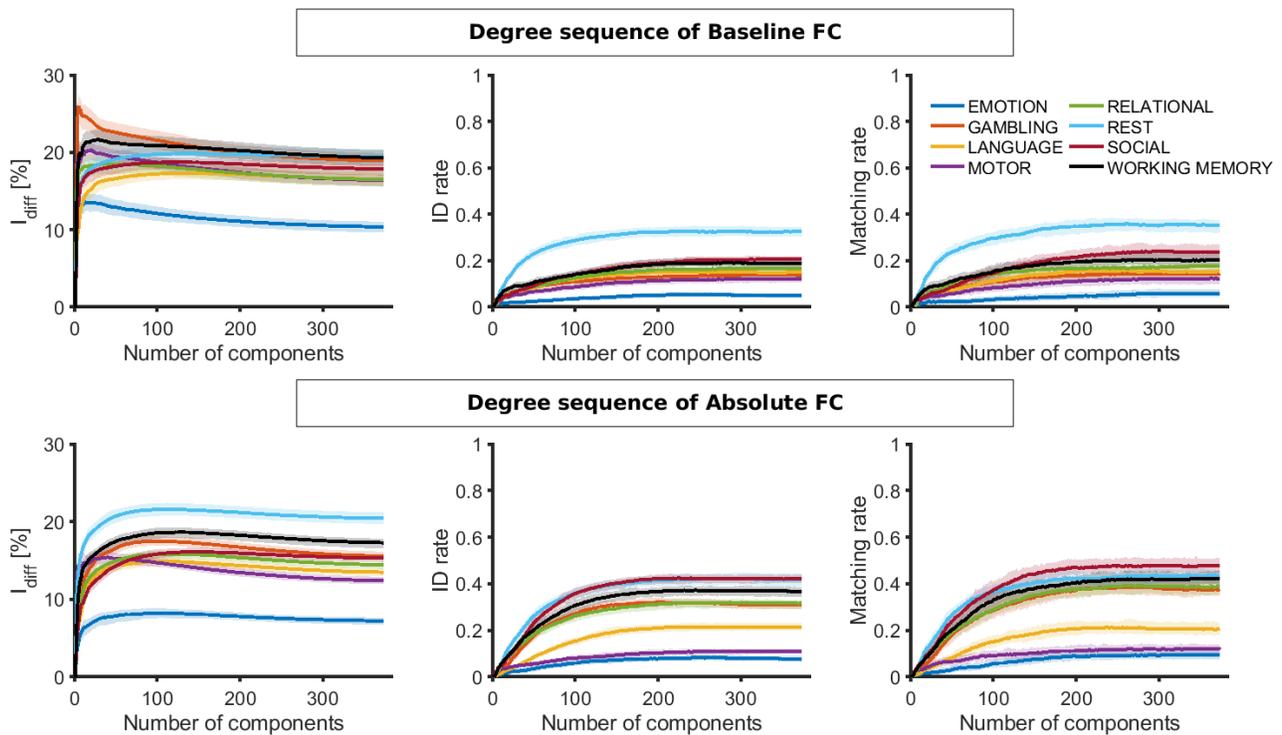

**Fig. S3. Fingerprinting the weighted degree sequence.** We applied the PCA decomposition-reconstruction procedure [Amico and Goñi (2018)] to a data matrix whose columns are the weighted degree sequence of both test and retest FCs, instead of using the vectorized upper triangular part of FC matrices as done in the main analysis. Solid lines represent the median value across 100 random subsamples of the database and shaded areas correspond to the inter-percentile range (2.5 and 97.5 percentiles). Top row: the weighted degree sequence is computed from the baseline FC. Bottom row: the weighted degree sequence is computed from the absolute FC. We computed the three metrics used in the main manuscript: differential identifiability (left), identification rate (middle) and matching rate (right).

**Table T1. Number of frames for each fMRI condition.**

| Condition | Number of frames |
|---|---|
| EMOTION | 176 |
| GAMBLING | 253 |
| LANGUAGE | 316 |
| MOTOR | 284 |
| RELATIONAL | 232 |
| RESTING-STATE | 1200 |
| SOCIAL | 274 |
| WORKING MEMORY | 405 |